\documentclass[%
 reprint,
 amsmath,amssymb,
 aps,
]{revtex4-1}

\usepackage{graphicx}
\usepackage{dcolumn}
\usepackage{bm}

\usepackage{xcolor}

\begin{document}

\preprint{APS/123-QED}

\title{Morphing Continuum Theory for Turbulence:\\ Theory, Computation and Visualization}

\author{James Chen}
 \email{Email: jmchen@ksu.edu}
\affiliation{%
 Multiscale Computational Physics Lab\\
 Department of Mechanical and Nuclear Engineering\\
 Kansas State University
}%

\date{\today}

\begin{abstract}
A high order morphing continuum theory (MCT) is introduced to model highly compressible turbulence. The theory is formulated under the rigorous framework of rational continuum mechanics. A set of linear constitutive equations and balance laws are deduced and presented from the Coleman-Noll procedure and Onsager's reciprocal relations. The governing equations are then arranged in conservation form and solved through the finite volume method with a second order Lax-Friedrichs scheme for shock preservation. A numerical example of transonic flow over a three-dimensional bump is presented using MCT and the finite volume method. The comparison shows that MCT-based DNS provides a better prediction than NS-based DNS with less than 10\% of the mesh number when compared with experiments. A MCT-based and frame-indifferent Q-criterion is also derived to show the coherent eddy structure of the downstream turbulence in the numerical example. It should be emphasized that unlike the NS-based G-criterion, the MCT-based Q-criterion is objective without the limitation of Galilean-invariance.
\end{abstract}

\maketitle


\section{\label{sect:Intro}Introduction}
Navier-Stokes (NS) equations have been extensively used  to study flow physics for several decades. Other than deriving the balance laws from the Reynolds Transport Theorem and the three-dimensional Leibniz Theorem common seen in undergraduate textbooks \cite{Batchelor1967}, these laws can also be derived independently from rational continuum mechanics (RCM) \cite{Eringen1971,Eringen1980,TruesdellRajagopal1999,Truesdell1966} or Boltzmann's kinetic theory \cite{Boltzmann1964,Huang1963,Ferziger1972}. or Classical Irreversible Thermodynamics \cite{Gyarmati1961_1, Gyarmati1961_2, Li1962,ColemanNoll1963, Chen2013, MullerRuggeri1991}. This  procedure  pairs  independent variables and response functions as thermodynamic forces and fluxes, i.e. a thermodynamic conjugate \cite{Gyarmati1961_1,Gyarmati1961_2,Li1962}. The Helmholtz free energy is  expanded with thermodynamic forces and the thermodynamic flux is then found as the derivative of the Helmholtz free energy with respect to the corresponding thermodynamic force. The  derivation process for  linear constitutive equations  was understood as the Onsager{\textendash}Casimir relations \cite{Onsager1931-1, Onsager1931-2, DegrootMazur1962}. Nevertheless, though the framework is mathematically rigorous and theoretically sounding, understanding the physics represented by the material constants in those equations heavily relies on experimental observation and measurements.

At the same time, kinetic theory  approximates the  gas atoms as points and model the interaction as collisions. Boltzmann derived a distribution function for equilibrium states with H-theorem and introduced a conservation equation with a collision integral \cite{Huang1963,Boltzmann1964}. With the proper definitions for kinetic variables, e.g. mass, linear momentum and energy, the conservations equations leads to the balance laws \cite{Ferziger1972}. However, systems are  rarely in equilibrium. As a result, the distribution  function varies with the interactions between gas atoms through the collision integral.  If the distribution is assumed to linearly deviate from the Boltzmann distribution,  the  balance law of linear momentum leads to the celebrated NS equations \cite{Huang1963}. Since  kinetic theory is a physics-based approach, the physical meanings of material constants are explained through the inter-particle collisions. For example, Maxwell showed that the viscosity is independent of the density for a given temperature through kinetic theory and later verified this fact with experiments \cite{Huang1963}. For dilute gases, kinetic theory also shows a linear relation for the ratio of thermal conductivity to the product of the viscosity and specific heat.  Though kinetic theory provides a detailed insights for the governing equations and material constants, Truesdell raised a concern that all equations derived from kinetic theory only contain a subset of those from rational continuum theory and 
claimed the validity of the continuum equations are beyond rarefied gases \cite{Truesdell1984}.

Regardless of different theoretical origins, NS equations have been the core of the fluid dynamics research ranging from turbulence to vortex-dominated flows for decades.  Nevertheless, assumptions made in NS equations should always be kept in mind. Rational continuum mechanics assumes a continuum homogenizing volumeless points. On the other hand, kinetic theory approximates continuous medium populated with monoatomic gases.  This approximation results in a compromise of relying on an orbital angular velocity, i.e. vorticity, to describe rotational motions in fluids. In other words, vorticity or vorticity-based methods have been employed to describe the rotational eddies in turbulent flows,  coherent vorticies in dynamic stall or body-vortex interaction in biomimetics. However, vorticity-based approaches are only Galilean invariant and present inconsistent results for vortex visualization from Q-criterion and $\lambda_2$ method in rotating flows \cite{Haller2005}. Therefore, Speziale and  Haller have been emphasizing the importance of objectivity or frame-indifference for vortex descriptions and turbulence models \cite{Speziale1979,Speziale1987,SpezialeSarkarGatski1991,Haller2005}. Further, revealing the detailed eddy structures with vorticity-based method requires extremely fine mesh for numerical differentiation. These arbitrarily fine meshes cause heavy computational burden and make numerical simulation impractical even with the best computational power available today \cite{ZhongWang2012}.

The Cosserat Brothers initiated the concept of a morphing continuum theory (MCT), i.e. a continuous space containing inner structures \cite{Cosserat1909}. Later, Eringen formulated a class of morphing continuum allowing rotation at subscale under the framework of rational continuum mechanics and classical irreversible thermodynamics \cite{Eringen1999}. Independently, 
Grad introduced a concept of a continuum with arbitrary number of internal degrees of freedom and proposed a first order differential equation for spins \cite{Grad1952}. De Groot and Mazur extended Grad's formulation and proposed a balance law for the internal spins \cite{DegrootMazur1962}. Snider and Lewchuk later completed Grad's formulations with irreversible thermodynamics \cite{SniderLewchuk1967}. The theoretical studies independently initiated by Eringen and Grad were found to be identical.

Similar to the works done for NS equations with statistical mechanics and kinetic theory, She and Sather relied on the Chapman-Enskog method to derive a kinetic theory for molecules with arbitrary internal degrees of freedom \cite{SheSather1967}. Brau evaluated several different collision processes under She and Sather's formulations \cite{Brau1967}. Curtiss later integrated these studies and officially introduced kinetic theory for molecular gases \cite{Curtiss1992}.

To the author's limited knowledge, the detailed comparison between irreversible thermodynamics/rational continuum mechanics and statistical kinetic theory was first presented in the book, {\textit{Rational Extended Thermodynamics}}, by M{\" u}ller and Ruggeri\cite{MullerRuggeri1991}. Rational Extended Thermodynamics has been used to derive governing equations for shock wave structure, light scattering, radiation, relativistic mechanics, phonons and metal electrons \cite{MullerRuggeri1991}. For monoatomic gases, the phenomenological equations derived from kinetic theory are found to be identical to those from thermodynamics of irreversible process. Motivated by these early studies, Chen recently proved that the inviscid version of the MCT governing equations from rational continuum mechanics/irreversible thermodynamics are identical to the balance laws at equilibrium from Curtiss' molecular kinetic theory \cite{Chen2016, ChenNew}. This study unifies both formulations in kinetic theory and rational continuum mechanics for fluid system with internal spins, and derives the Boltzmann-Curtiss distribution from a quantum statistics perspective \cite{Chen2016,ChenNew}.

Since its introduction , MCT has been used to study the flow physics with significant internal spins, eg. turbulence \cite{KirwanNewman1969, Liu1970, EringenChang1970, Peddieson1972, Ahmadi1975, Ahmadi1981, BrutyanKrapivsky1992}.  Ahmadi extended the work of Liu \cite{Liu1970} to construct a statistical theory for turbulence via a functional approach \cite{Ahmadi1981}. Peddieson was the first one proposing  MCT dimensionless parameters to characterize the wall shear layers in the boundary layer turbulence \cite{Peddieson1972}. More recently, Mehrabian and Atefi compared the analytical solution of plane Poiseulle flow in MCT with the experimental velocity profiles, both laminar and turbulent \cite{MehrabianAtefi2008}. Alizadeth et. al. reformulated MCT and studied the turbulent plane Couette flow with slip. The simulation results agree extremely well with the experiments \cite{AlizedethSilberNejad2011}. However, all the aforementioned studies are limited in the analytical solutions for incompressible flows. More recently, researchers have been focusing on developing numerical methods for MCT in both incompressible and compressible flows \cite{ChenLiangLee2010, ChenLiangLee2012, MujakovicCrnjavic2016, DrazicMujakovicCrnjavic2017}.

With the rapid developments on the numerical solvers, a series of studies were published on incompressible and compressible turbulence and their statistical characteristics \cite{WonnellChen2016, WonnellChen2017, CheikhChen2017, WonnellChen2017-2}.  Cheikh and Chen validated that MCT is capable of predicting the velocity profile over a compression ramp in a supersonic turbulence \cite{CheikhChen2017}. However, MCT is still relatively new to turbulence community. Tools for turbulence analysis and visualization are at their infancy. Therefore, this study will summarize the derivation of MCT, its connection to turbulence, and an objective tool for visualizing coherent eddy structures. The derivation of MCT is briefly reviewed in Section \ref{sect:theory}.  The MCT governing equations will be rearranged in the conservation forms for numerical methods implementations, e.g. finite difference method, finite volume method  and others, in Section \ref{sect:numerical}. In Section \ref{sect:example}, a numerical example of a transonic flow over a three-dimensional bump will be briefly presented. An objective Q-critirion with MCT will be derived and discussed in details for eddy and vortex visualizations.  Section \ref{sect:conclusion} concludes the highlights and remarks of this study.

\section{Morphing Continuum Theory}\label{sect:theory}
A morphing continuum is a collection of continuously distributed,
oriented, finite-size subscale structures that allows rotations.  A material point $P$
in the reference frame is identified by a position and three directors
attached to the material point.

The motion, at time $t$, carries the finite-size subscale structure to a spatial
point and rotates the three directors to a new orientation.  Thus, such a motion
can be understood as the motion of a liquid molecule or an eddy approximated as a rigid body.
MCT possesses not only translational velocity, but also
self-spinning gyration on its own axis.  These motions and their inverse
motions for the morphing continuum can be 
described as \cite{Eringen1999}
\begin{align}   \label{eq:director}
   x_k&=x_k(X_K, t) \quad & X_K&=X_K(x_k, t)\nonumber \\
   \xi_k&=\chi_{kK}(X_K, t)\Xi_K	\quad & \Xi_K&=\bar{\chi}_{Kk}\xi_k\\
   K&=1, 2, 3 \quad &  k&=1, 2, 3\nonumber
\end{align}
and
\begin{align}\label{eq:inverse}
   \chi_{kK}\chi_{lK}=\delta_{kl} \qquad \bar{\chi}_{Kk}\bar{\chi}_{Lk}=\delta_{KL}.
\end{align}
where the lowercase index is for the Eulerian coordinate while the uppercase index for the Lagrangian coordinate.

It is straightforward to prove that
\begin{align}
    \chi_{kK}=\bar{\chi}_{Kk}.
\end{align}
Consequently, the righthand side of eq.~\ref{eq:inverse} becomes
\begin{align}
    \chi_{kK}\chi_{kL}=\delta_{KL} .
\end{align}
Here and throughout, an index followed by a comma denotes a partial derivative, e.g.,
\begin{align}
    x_{k,K}=\frac{\partial x_k}{\partial X_K}  \qquad\mbox{and}\qquad
      X_{K,k}=\frac{\partial X_K}{\partial x_k}.
\end{align}

For fluid flow, deformation-rate tensors are used to characterize the
viscous resistance.  Deformation-rate tensors may be deduced by
calculating the material time derivative of the spatial deformation
tensors. For a morphing continuum, two objective deformation-rate tensors
can be derived as
\begin{equation}
   a_{kl}=v_{l,k}+e_{lkm}\omega_m
    \qquad\mbox{and}\qquad   b_{kl}=\omega_{k,l},
\end{equation}
where $v_k$ is the velocity vector and
$\omega_k$ is the self-spinning gyration vector.  The fluid or flow inner structure possesses two
types of motion, translational velocity ($v_k$), found
by solving the MCT linear momentum equation, and spinning gyration
($\omega_k$) found by solving the MCT angular momentum
equation.  In the classical Navier-Stokes equations, the
translational velocity can be directly solved from the balance law of
linear momentum.  To investigate the effect of the rotational motion of
the subscale structure, one must use the velocity field and take the angular
velocity to be one-half of the vorticity i.e.,
$\frac{1}{2}e_{ijk}v_{j,i}$.  This approximation in the
Navier-Stokes equations limits not only predicting the flow physics
involving spinning, but also fails to represent the
interaction between translation and spinning \cite{SniderLewchuk1967}.
In addition, highly refined meshes are also needed in order to obtain
high resolution vorticity fields. This arbitrary mesh requirement makes
numerical simulations in realistic environments impractical \cite{ZhongWang2012}.
On the other hand, MCT provides both the self-spinning motion and the relative rotation,
e.g. vorticity. The arbitrarily fine meshes are no longer required since part of the information
on rotational motions can be directly obtained from self-spinning gyration.

\subsection{Balance Laws}
Thermodynamic balance laws for morphing continuum theory include
(1) mass; (2) linear momentum; (3)
angular momentum; (4) energy; and (5)
the Clausius-Duhem inequality.  All five can be expressed as follows:

{\textit {Conservation of mass}}
\begin{align}
    \frac{\partial\rho}{\partial t}+(\rho v_i)_{,i}=0
\label{eq:MCT+Mass}
\end{align}

{\textit {{Balance of linear momentum}}
\begin{align}
    t_{lk,l}+\rho(f_k-\dot{v}_k)=0
\label{eq:MCT+LM}
\end{align}

{\textit {{Balance of angular momentum}}
\begin{align}
    m_{lk,l}+e_{ijk}t_{ij}+\rho i_{km}(l_m-\dot{\omega}_m)=0
\label{eq:MCT+AM}
\end{align}

{\textit {{Balance of energy}}
\begin{align}
    \rho\dot{e}-t_{kl}a_{kl}-m_{kl}b_{lk}+q_{k,k}=0
\label{eq:MCT+E}
\end{align}

{\textit {Clausius-Duhem inequality}}
\begin{align}
    \rho(\dot{\psi}+\eta\dot{\theta})+t_{kl}a_{kl}+m_{kl}b_{lk}-\frac{q_k}{\theta}\theta_{,k}\geq0
\label{eq:CDI}
\end{align}
where $\rho$ is mass density, $i_{km}$ the microinertia for the shape of the
microstructure, $f_k$ the body force density, $l_m$ the body moment
density, $e$ the internal energy density, $\eta$ the entropy density,
$\psi=e-\theta\eta$ the Helmholtz free energy, $t_{lk}$ the Cauchy stress,
$m_{lk}$ the  moment of stress, and $q_k$ the heat flux.  It is worthwhile to
mention that $t_{lk}$, $m_{lk}$ and $q_k$ are the constitutive equations
for the morphing continuum theory and can be derived from the
Clausius-Duhem inequality (see eq.~\ref{eq:CDI}) through the Coleman-Noll
procedure \cite{ChenLeeLiang2011, ColemanNoll1963}.

The concept of subscale inertia,
$i_{km}\equiv{\int\rho'\xi_k\xi_mdv'}/{\int\rho'dv'}\equiv\langle\xi_k\xi_m\rangle$,
is similar to the moment of inertia in rigid body rotation and measures the
resistance of the subscale structure to changes to its rotation.  It can be
further expressed as
\begin{equation}
   j_{km}=i_{pp}\delta_{km}-i_{km} \qquad\mbox{where}\qquad
   j\equiv\frac{1}{3}j_{pp}.
\end{equation}

The volume $v'$ refers to the volume of the subscale structure.
If the subscale structure is assumed to be a rigid sphere with a radius
$d$ and a constant density $\rho$, the subscale inertia can be computed as
$j=\frac{2}{5}d^2$. This result shows the subscale inertia for a sphere
is the moment of inertia of a sphere divided by its mass.
The experimental data of Lagrangian velocities of a trace particle can
be used to determine the geometry of the subscale structure
\cite{ChenLeeLiang2011, MordantMetzMichelPinton2001}. 
{The new degrees of freedom, gyration,
in MCT can also be directly compared with the direct experimental measurement of vorticity \cite{RyabtsevPouya2016}.}}

\subsection{Constitutive Equations}
There are multiple different definitions for fluids, including (1)
fluids do not have a preferred shape \cite{Batchelor1967}, and (2), fluids
cannot withstand shearing forces, however small, without sustained
motion \cite{Panton2013}.  Nevertheless, all these definitions describe
the physics of fluid flow, and yet provide little help in mathematically
formulating a continuum theory for fluids.  In rational continuum
mechanics, Eringen formally defined fluids by saying that
``a body is called fluid if every configuration of
the body leaving density unchanged can be taken as the reference
configuration"\cite{Eringen1980}. This definition implies $x_{k,K}\rightarrow\delta_{kK}$ and
$\chi_{kK}\rightarrow\delta_{kK}$ where $\delta_{kK}$ is the shifter,
the directional cosine between the current configuration and reference
configuration.  

{Objectivity is followed throughout the derivation of constitutive equations}.}
 The axiom of objectivity, or frame-indifference, states that the constitutive
 equations must be form-invariant with respect to rigid body motions of the spatial frame
of reference \cite{Eringen1971, Eringen1980,Truesdell1966,TruesdellRajagopal1999}.

The state of fluids in morphing continuum theory is expressed by the
characterization of the response functions $\mathbf{Y}=\{\psi, \eta,
t_{kl}, m_{kl}, q_k\}$ as functions of a set of independent variables
$\mathbf{Z}=\{\rho^{-1}, \theta, \theta_{,k}, a_{kl}, b_{kl}\}$. At the outset the
constitutive relations are written as
$\mathbf{Y}=\mathbf{Y}(\mathbf{Z})$ \cite{Eringen2001}.

The Clausius-Duhem inequality of eq.~\ref{eq:CDI}, also known as the
thermodynamic second law, is a statement concerning the irreversibility
of natural processes, especially when energy dissipation is involved.
Feynman et.\ al.\ stated ``so we see that a substance must be limited in a
certain way; one cannot make up anything he wants; ...  This [entropy]
principle, this limitation, is the only rule that comes out of
thermodynamics"  \cite{FeymanLighttonSand1970}.  After the Coleman-Noll
procedure, i.e., combining the inequality with the response function and
the independent variables, eq.~\ref{eq:CDI} reduces to
\begin{align}
   t_{kl}^d a_{kl}+m_{kl} b_{lk}-\frac{q_k \theta_{,k}}{\theta}\geq0 .
   \label{eq:FinalCDI}
\end{align}
{The current formulation relying on the Coleman-Noll procedure
only provides local entropy increase and the conditions for first order weakly nonlocal
state.}

In eq.~\ref{eq:FinalCDI}, there are three pairs of thermodynamic conjugates, ($t_{kl}^d$,
$a_{kl}$), ($m_{kl}$, $b_{lk}$), and
($\frac{q_k}{\theta}$,$\theta_{,k}$) that contribute to the
irreversibility of the material.  A set of the thermodynamic fluxes
$\mathbf{J}$ is defined as $\mathbf{J}=\{t_{kl}^d, m_{kl},
\frac{q_k}{\theta}\}$ and are functions of a set of the thermodynamic forces
($\mathbf{Z}^D$) and other independent variables ($\mathbf{Z}^R$),
$\mathbf{Z}=\{\mathbf{Z}^R; \mathbf{Z}^D\}=\{\rho^{-1}, \theta; a_{kl},
b_{lk}, \theta_{,k}\}$.  With these sets of thermodynamic fluxes and
thermodynamic forces, the Clausius-Duhem inequality can be rewritten as
\begin{align}
   \mathbf{J}(\mathbf{Z}^R; \mathbf{Z}^D)\cdot\mathbf{Z}^D\geq0 .
\end{align}

Onsager and {{others}} proposed that the thermodynamic fluxes can be
obtained by the general dissipative function \cite{Chen2013, Edelen1972,
Onsager1931-1, Onsager1931-2, Gyarmati1961_1, Gyarmati1961_2, Li1962}
\begin{align}
   \mathbf{J}=\frac{\partial
   \Psi(\mathbf{Z}^R,\mathbf{Z}^D)}{\partial\mathbf{Z}^D}+\mathbf{U} ,
\end{align}
where the vector-valued function $\mathbf{U}$ is the constitutive
residual with $\mathbf{Z}^D\cdot\mathbf{U}=0$.  This result indicates
that $\mathbf{U}$ does not contribute to the dissipative or entropy
production.  For simplicity, one can further set $\mathbf{U}=0$.

To determine thermodynamic fluxes for a fluid using the derivative of
$\Psi$ with respect to the thermodynamic forces $\mathbf{Z}^D$, $\Psi$ needs to be
invariant under superimposed rigid body motion, i.e., the dissipative
function $\Psi$ must satisfy the axiom of objectivity \cite{Chen2013}.
Hence, $\Psi$ is an isotropic function of scalar and can be expressed by
Wang's representation theorem \cite{Wang1969, Wang1970} as
\begin{align}
   &\Psi\{\mathbf{Z}^R, \mathbf{Z}^D\}=\Psi\{I_1, I_2, I_3, ..., I_n\}\nonumber\\
   \mbox{and}\qquad & \nonumber\\
   &\mathbf{J}=\frac{\partial\Psi}{\partial\mathbf{Z}^D}=
     \sum_{i=1}^{n}\frac{\partial\Psi}{\partial I_i}
     \frac{\partial I_i}{\partial \mathbf{Z}^D} .
\end{align}
It should be noted here that $b_{lk}$ and $m_{kl}$ are pseudo-tensors
while the rest, including $\theta_{,k}$, $t_{kl}^d$, $q_k$ and
$a_{kl}$ are normal tensors \cite{ChenLeeLiang2011}.  Considering the
mixing of pseudo-tensors and normal tensors in $\mathbf{Z}^R$ and
$\mathbf{Z}^D$ for linear constitutive equations, the set of invariants
includes $I_1=a_{(ii)}$, $I_2=a_{(ij)}a_{(ji)}$, $I_3=b_{(ij)}b_{(ji)}$,
$I_4=\theta_{,k}\theta_{,k}$, $I_5=a_{[ij]}a_{[ji]}$, $I_6=b_{[ij]}b_{[ji]}$,
and $I_7=e_{ijk}b_{ij}\theta_k$ .
Here (...) refers to the symmetric part, [...] indicates the
anti-symmetric part and $e_{ijk}$ is the permutation symbol.  Hence, the
thermodynamic fluxes can be further derived as
\begin{align}
   t_{kl}^d& =t_{(kl)}^d+t_{[kl]}^d\nonumber\\
   & =\frac{\partial\Psi}{\partial
   a_{(kl)}}+\frac{\partial\Psi}{\partial
   a_{[kl]}}\nonumber\\
   & =\frac{\partial\Psi}{\partial
   I_1}\delta_{kl}+\frac{\partial\Psi}{\partial
   I_2}a_{(kl)}+\frac{\partial\Psi}{\partial I_5}a_{[kl]}\nonumber \\
   & = \lambda a_{mm}\delta_{kl}+2\mu a_{(kl)}+\kappa (a_{(kl)}+a_{[kl]})\nonumber \\
   m_{kl}& =m_{(kl)}+m_{[kl]}\nonumber\\
   & =\frac{\partial\Psi}{\partial
   b_{(kl)}}+\frac{\partial\Psi}{\partial
   b_{[kl]}}\nonumber\\
   & =\frac{\partial\Psi}{\partial
   I_3}b_{(kl)}+\frac{\partial\Psi}{\partial
   I_6}b_{[kl]}+\frac{\partial\Psi}{\partial
   I_7}e_{klm}\theta_{,m}\nonumber \\
   & =\alpha b_{mm}\delta_{kl} + \frac{1}{2}(\beta+\gamma)b_{(kl)}+\frac{1}{2}(\beta-\gamma)b_{[kl]} + \frac{\alpha_T}{\theta}e_{klm}\theta_{,m}\nonumber\\
   q_k& =\frac{\partial\Psi}{\partial\theta_{,k}}\nonumber\\
   & =\frac{\partial\Psi}{\partial
   I_4}\theta_{,k}+\frac{\partial\Psi}{\partial I_7}e_{ijk}b_{ij}\nonumber \\
   & =K\theta_{,m} + \frac{\alpha_T}{\theta}e_{klm}b_{[kl]}
\end{align}
These equations can also be put into matrix form as
\begin{align}
&\quad\begin{bmatrix}
   t_{(kl)}^d \\
   t_{[kl]}^d \\
   m_{(kl)} \\
   m_{[kl]} \\
   q_k
\end{bmatrix}
\\
&= \begin{bmatrix}
   \lambda\delta_{kl}+2\mu+\kappa       &0      &0      &0      &0\\
   0    &\kappa &0      &0      &0\\
   0    &0      &\alpha\delta_{kl}+\frac{1}{2}(\beta+\gamma) &0      &0\\
   0    &0      &0      &\frac{1}{2}(\beta-\gamma)   &\frac{\alpha_Te_{klm}}{\theta}\\
   0    &0      &0      &\frac{\alpha_Te_{klm}}{\theta} &K
\end{bmatrix}\nonumber\\
& \begin{bmatrix}
   a_{(kl)}\\
   a_{[kl]}\\
   b_{(kl)}\\
   b_{[kl]}\\
  \theta_{,m}
\end{bmatrix}
\nonumber\\
&= \begin{bmatrix}
   \lambda\delta_{kl}+2\mu+\kappa       &0      &0      &0      &0\\
   0    &\kappa &0      &0      &0\\
   0    &0      &\alpha\delta_{kl}+\frac{1}{2}(\beta+\gamma) &0      &0\\
   0    &0      &0      &\frac{1}{2}(\beta-\gamma)   &\frac{\alpha_Te_{klm}}{\theta}\\
   0    &0      &0      &\frac{\alpha_Te_{klm}}{\theta} &K
\end{bmatrix}\nonumber\\
& \begin{bmatrix}
   \frac{1}{2}(v_{k,l}+v_{l,k})\\
   \frac{1}{2}(v_{l,k}-v_{k,l}+2e_{lkm}\omega_m)\\
   \frac{1}{2}(\omega_{k,l}+\omega_{l,k})\\
   \frac{1}{2}(\omega_{k,l}-\omega_{l,k})\\
  \theta_{,m}
\end{bmatrix} .
\label{eq:ThermoMatrix}
\end{align}
Notice the symmetry of this thermodynamic matrix.
Equations \ref{eq:ThermoMatrix} connect the thermodynamic fluxes and
the thermodynamic forces, and can be referred to as Onsager's reciprocal
relations derived in 1931 \cite{Onsager1931-1, Onsager1931-2} leading
to his Nobel Prize in Chemistry in 1968.
{{It should be noted that the reciprocity is the condition for the existence
of the dissipative potential \cite{Silhavy1997,BerezovskiVan2017}.}}
 With further algebraic
manipulation, the linear constitutive equations for the morphing continuum
are
\begin{align}
  t_{lk}& =-p\delta_{kl}+\lambda v_{m,m}\delta_{kl}+\mu(v_{k,l}+
   v_{l,k})+\kappa
   (v_{k,l}+e_{klm}\omega_m)\nonumber\\
   m_{lk}& =\frac{\alpha_T}{\theta}e_{lkm}\theta_{,m}+\alpha\omega_{m,m}
   \delta_{kl}+\beta\omega_{l,k}+\gamma\omega_{k,l}\nonumber\\
   q_k& =\frac{\alpha_T}{\theta}e_{lkm}\omega_{k,l}+K\theta_{,m}
\label{eq:MCT+Con}
\end{align}

where $\mu$ is the viscosity, $\lambda$ is the secondary viscosity, $\kappa$ is the subscale viscosity, $\gamma$ is the subscale diffusivity and $\alpha$ \& $\beta$ are related to the compressibility of the subscale structure. Inserting eq. \ref{eq:MCT+Con} to all the balance laws, eqs. \ref{eq:MCT+Mass}-\ref{eq:MCT+E},
omitting body force and adopting a spherical subscale structure,
the MCT governing equations can be rewritten as

{\textit {Conservation of mass}}
\begin{align}
    \frac{\partial\rho}{\partial t}+(\rho v_i)_{,i}=0
\label{eq:MCT+Mass_G}
\end{align}

{\textit {Balance of linear momentum}}
\begin{align}
   & \rho(\frac{\partial v_k}{\partial t} + v_iv_{k,i})\nonumber\\
 &= -p_k + (\lambda+\mu)v_{m,mk} + (\mu+\kappa) v_{k,ll}+\kappa e_{kij}\omega_{j,i}
\label{eq:MCT+LM_G}
\end{align}

{\textit {Balance of angular momentum}}
\begin{align}
& \rho j (\frac{\partial \omega_m}{\partial t} + v_i\omega_{m,i})\nonumber\\
 &= (\alpha+\beta)\omega_{m,mk} + \gamma\omega_{m,ll} + \kappa (e_{mnk}v_{k,n} - 2\omega_m)
\label{eq:MCT+AM_G}
\end{align}

{\textit {Balance of energy}}
\begin{align}
& \rho (\frac{\partial e}{\partial t} + v_ie_{,i})\nonumber\\
&=[-p\delta_{kl}+\lambda v_{m,m}\delta_{kl}+ \mu(v_{k,l}+
   v_{l,k})+\kappa
   (v_{l,k}+e_{lkm}\omega_m)] \nonumber\\
   &\qquad\qquad\qquad\qquad\qquad\qquad\qquad\qquad\qquad(v_{l,k}+e_{lkm}\omega_m)\nonumber\\
   &+ (\frac{\alpha_T}{\theta}e_{klm}\theta_{,m}+\alpha\omega_{m,m}
   \delta_{kl}+\beta\omega_{k,l}+\gamma\omega_{l,k})\omega_{k,l}\nonumber\\
   &+K\theta_{,mm}
\label{eq:MCT+E_G}
\end{align}

\section{Numerical Methods}\label{sect:numerical}
Equations \ref{eq:MCT+Mass_G} - \ref{eq:MCT+E_G} can be directly discretized and solved with the classical finite difference method \cite{ChenLiangLee2010}; however, in order to adopt modern numerical schemes, such as finite volume method \cite{CheikhChen2017,Patankar1984}, spectral difference method \cite{ChenLiangLee2012, LiuVinokurWang2006, WangLiuMayJameson2007}, spectral volume method \cite{Wang2002, LiuVinokurWang2006_2} and others \cite{NishikawaLiu2017}, the governing equations should be cast into the conservation forms. Chen et. al. formulated the conservation form for MCT as \cite{ChenLiangLee2012}
\begin{align}
&\frac{\partial\rho}{\partial t}+\nabla\cdot(\rho\vec{v})=0\label{eq:MCT+M+C}\\
&\frac{\partial\rho\vec{v}}{\partial t}+\nabla\cdot(\vec{v}\otimes\rho\vec{v})=\nonumber\\
&\quad -\nabla p + (\lambda+\mu)\nabla\nabla\cdot\vec{v} + (\mu+\kappa)\nabla^2\vec{v} + \kappa\nabla\times\vec{v}\label{eq:MCT+LM+C}\\
&\frac{\partial\rho j\vec{\omega}}{\partial t}+\nabla\cdot(\vec{v}\otimes\rho j\vec{\omega})= \nonumber\\
&\qquad\qquad(\alpha+\beta)\nabla\nabla\cdot\vec{\omega} + \gamma\nabla^2\vec{\omega} + \kappa(\nabla\times\vec{v} - 2\vec{\omega})\label{eq:MCT+AM+C}\\
&\frac{\partial \rho E}{\partial t} + \nabla\cdot(\vec{v}\rho E) = \nonumber \\
&\qquad\qquad\qquad\qquad\nabla\cdot(\mathbf{t}\cdot\vec{v}) + \nabla\cdot(\mathbf{m}\cdot\vec{\omega})-\nabla\cdot\vec{q}\label{eq:MCT+E+C}
\end{align}
where $\mathbf{t}$ is the MCT Cauchy stress, $\mathbf{m}$ is the MCT moment stress and $\vec{q}$ is the MCT heat flux (cf. eq.\ref{eq:MCT+Con}). In addition, $E$ is the MCT total energy defined as the sum of inernal energy density, translational kinetic energy density and rotational kinetic energy density as $E=e+\frac{1}{2}(\vec{v}\cdot\vec{v} + j\vec{\omega}\cdot\vec{\omega})$.

Several researchers have started focusing on numerical methods for MCT \cite{ChenLiangLee2010, ChenLiangLee2012, DrazicMujakovicCrnjavic2017, MujakovicCrnjavic2016, MujakovicCrnjaricZic2015}. More specifically, three different numerical schemes are introduced in the past few years: (1) finite difference method with second order temporal and spatial accuracy for incompressible flows \cite{ChenLiangLee2010}; (2) finite volume method with second order shock preserving scheme for compressible flows \cite{CheikhChen2017}; and (3) high order spectral difference method for compressible flows \cite{ChenLiangLee2012}. In this study, the finite volume method with second order shock preserving scheme is used and summarized as follows.

The finite volume method  directly implements a numerical procedure solving the conservation forms of governing equations. The general form over a control volume can be expressed as
\begin{align}
\frac{\partial\Psi}{\partial t}+\nabla\cdot(\vec{v}\Psi)-\nabla\cdot(\Gamma\nabla\Psi)=S_{\Psi}
\end{align}
where  $\Psi$ is any physical quantity, $\frac{\partial\Psi}{\partial t}$ is the unsteady term, $\nabla\cdot(\vec{v}\Psi)$ is the convective term,  $\nabla\cdot(\Gamma\nabla\Psi)$ is the diffusion term and $S_{\Psi}$ is the source term.

The diffusion term is discretized by the central difference method and Green-Gauss theorem, i.e.
\begin{align}
\int_V\nabla\cdot(\Gamma\nabla\Psi)dV&=\oint_{\vec{A}}\Gamma\nabla\Psi\cdot d\vec{A}\nonumber \\
&\approx\sum_f\Gamma_f\vec{A}_f\cdot\nabla\Psi_f
\end{align}
where  $V$ is the control volume, $\vec{A}$ is the enclosed surfaces with normal vectors for the control volume and $f$ indicates each surfaces in the control volume for calculation.

The convective term is discretized  by the second order Lax-Friedrich flux splitting method by Kuganov, Noelle and Petrova (KNP) \cite{KuganovNoellePetrova2001}, i.e.
\begin{align}
\int_V \nabla\cdot(\vec{v}\Psi)dV&=\oint_{\vec{A}}\vec{v}\Psi\cdot d\vec{A}\nonumber \\
&\approx\sum_f\vec{A}_f\cdot\vec{v}_f\Psi_f\nonumber \\
&=\sum_f\phi_f\Psi_f
\end{align}
and
\begin{align}
\sum_f\phi_f\Psi_f=&\sum_f[\alpha\phi_{f+}\Psi_{f+}\nonumber\\
&+(1-\alpha)\phi_{f-}\Psi_{f-}+\omega_f(\Psi_{f-}-\Psi_{f+})]
\end{align}
where $\alpha$ is calculated based on the local speed of sound, ie.
\begin{align}
&\Psi_{f+}=\text{max}(c_{f+}|\vec{A}_f|+\phi_{f+},c_{f-}|\vec{A}_f|+\phi_{f-},0)\nonumber \\
&\Psi_{f-}=\text{max}(c_{f+}|\vec{A}_f|-\phi_{f+},c_{f-}|\vec{A}_f|-\phi_{f-},0)\nonumber \\
&\alpha=\frac{\Psi_{f+}}{\Psi_{f+}+\Psi_{f-}}\nonumber \\
&\omega_f=\alpha(1-\alpha)(\Psi_{f+}+\Psi_{f-})
\end{align}

The gradient and curl term is also discretized by the second order Lax-Friedrich flux splitting method, similar to the convective terms.
\begin{align}
\int_V\nabla\Psi dV&=\oint_{\vec{A}} \Psi d\vec{A}\nonumber \\
&=\sum_f \vec{A}_f \Psi_f
\end{align}
where the  KNP scheme further split the interpolation procedure into $f_+$ and $f_-$ direction, ie.
\begin{align}
\sum_f \vec{A}_f \Psi_f=\sum_f[\alpha\vec{A}\Psi_{f+}+(1-\alpha)\vec{A}\Psi_{f-}]
\end{align}

This second order generalized Lax-Friedrich flux was implemented in the NS framework. Cheikh and Chen further demonstrated that this approach also provides a second order accuracy in space for MCT \cite{CheikhChen2017}. If the discretized equations are solved in a explicit manner, the unsteady term can be marched with Runge-Kutta method of any order in temporal accuracy. In this study, a first order Euler method is chosen for convenience.

\section{Numerical Example - Transonic Flows over a 3D Bump}\label{sect:example}
There have been a significant amount of studies on turbulence simulation and analysis with morphing continuum theories since the 1970s \cite{Ahmadi1975,Ahmadi1981,BrutyanKrapivsky1992,EringenChang1970,Peddieson1972,KirwanNewman1969,MehrabianAtefi2008,WonnellChen2016,CheikhChen2017,WonnellChen2017,WonnellChen2017-2,AlizedethSilberNejad2011}. Most of the published efforts focused on the velocity profile with analytical means. With the assistance of the introduced finite volume method with shock preserving scheme in Sect. \ref{sect:numerical}, a transonic flow of Ma$=0.6$ over a 3D bump is simulated and compared with experimental measurements \cite{SimpsonLongByun2002} and {{NS-based direct numerical simulation (DNS)}} results \cite{CastagnaYaoYao2014}.

\subsection{MCT Simulation Compared with DNS and Experiments}
The inlet is specified with a turbulent boundary layer flow with a thickness of $\delta=39$ mm and the free stream velocity of $M_\infty=0.6$. The corresponding $Re_\delta*=500$ while $Re_\delta$ is 6.5 times of the $Re_\delta*$. The boundary layer flows over a three dimensional bump with the height of $H=78$ mm. The NS-based DNS (NS-DNS) study on this problem was reported and presented side by side with a comparable experiment in 2014 \cite{CastagnaYaoYao2014}. {{A MCT-based DNS (MCT-DNS) is performed following exactly the same setup reported in the NS-DNS study for the purpose of comparison. The pressure coefficient from NS-DNS (dashline) \cite{CastagnaYaoYao2014}, experiment (circle) \cite{SimpsonLongByun2002} and MCT-DNS (solid line) \cite{WonnellChen2017-2} is shown in Figure \ref{fig:Cp}. It can be seen that the NS-DNS only captures one experimental data point over the bum and is unable to clearly identify the normal shock. On the other hand, MCT-DNS captures most of the experimental data points over the normal shock and agrees better with the experiment.}} In addition, MCT also has a better prediction than NS on the pressure coefficient downstream. It should be mentioned that the deviation between the experiment and simulation is caused by the separation point in the flow. Neither NS nor MCT was able to predict the correct separation point. This inconsistency is due to the unknown channel surface properties, such as roughness, and the fluid properties. The fluid used in the experiment was kerosene while both NS and MCT simulation focused on the equivalence of dimensionless parameters comparable with experiments.
\begin{figure}[h!]
\includegraphics[height=2 in]{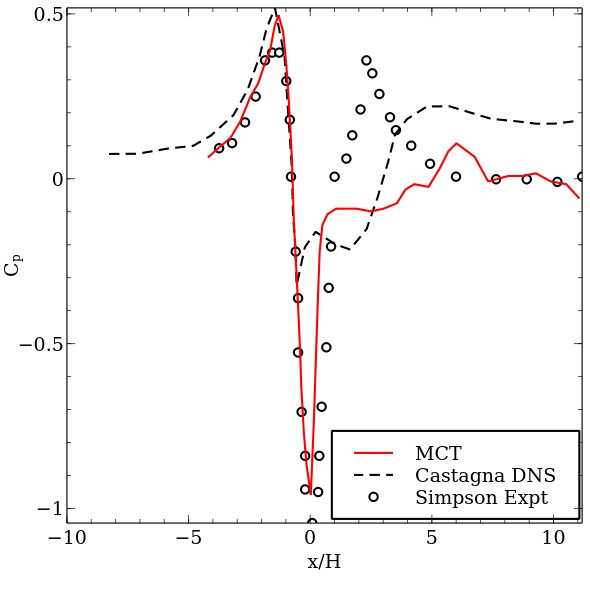}
\caption{Pressure coefficient comparison among experiment, DNS and MCT along the centerline on the bump}
\label{fig:Cp}
\end{figure}

The required computational resources between NS and MCT are also compared. The NS work adopts a high order finite difference method with a mesh number totaling $\sim$ 54M \cite{CastagnaYaoYao2014}. On the other hand, MCT was solved with a second order finite volume method with a shock preserving scheme and used a mesh number of $\sim$ 4.5M. With less than 10\% of the cell number required in DNS, MCT was able to have a better prediction on shock position and the pressure profile in the downstream turbulence. The multiscale nature of  MCT provides a rigorous framework coupling one level of motion for macroscale translation and another one for subscale eddy rotation. Therefore, there is no need for arbitrary fine mesh in capturing the subscale motion. In MCT, most of the subscale motions are captured by the additional degrees of freedom (gyration) at subscale. Part of these results were reported in AIAA Aviation 2017 \cite{WonnellChen2017-2}. {{Unlike the classical Reynolds-Averaging Numerical Simulation (RANS) or Large Eddy Simulation (LES), MCT does not require any turbulence models. The numerical solution of MCT is acquired in the same fashion of direct numerical simulation for NS equations. The flow at subscales are resolved by the additional degrees of freedom, gyration.}}

With a successful prediction from MCT, it is necessary to provide a tool to visualize the classical hairpin eddy structure in turbulence. However, the classical velocity-based criteria have been criticized on the inconsistency and limitation on being only Galilean invariant \cite{Haller2005}. The multiscale MCT can be further developed into a visualization tool with objectivity (or frame-indifference) and similar physical meanings provided by the classical criteria.

\subsection{Objective Description of Vortex Visualization}
{{Speziale devoted part of his career laying down the foundamentals of objectivity and investigated the requirment of objectivity over Galilean invariance for turbulence simulation  \cite{Speziale1979,Speziale1987,Speziale1989,Speziale1998,SpezialeSarkarGatski1991}}.} More recently, Haller showed the inconsistency of vortex identification with the classical velocity gradient-based  approaches and emphasized the importance of the objectivity or frame-indifference for vortex visualization \cite{Haller2005}. The classical Q-criterion under NS framework relies on the second invariant of the velocity gradient, eg. $2II_a=v_{i,i}v_{j,j}-v_{i,j}v_{j,i}=\Omega_{ij}\Omega_{ji}-S_{ij}S_{ji}$; where $v_{i,j}$ is velocity gradient, $S_{ij}=\frac{1}{2}(v_{i,j}+v_{j,i})$ and $\Omega_{ij}=\frac{1}{2}(v_{i,j}-v_{j,i})$. It has been proven that the symmetric part of velocity gradient, $S_{ij}$, is objective; however the antisymmetric part , $\Omega_{ij}$ is only Galilean invariant.

The objectivity or frame-indifference emphasizes the invariance between two reference frames. Let a rectangular frame, $M$, be in relative rigid motion with respect to another one, $M'$.  A point with  rectangular coordinate $x_k$ at time $t$ in $M$ will have another rectangular coordinate $x'_k$ at time $t'$ in $M'$.  Since the reference frames are  rigid motion with respect to each other, the motion between two frames can be described as $x'_k(t')=Q_{kl}(t)x_l(t)+b_k(t)$ where $Q_{kl}(t)$ is the rigid body rotation matrix between two frames and $b_{k}(t)$ is the translation between two frames. If the time derivative is performed on motion, it leads to $v'_k(t')=\dot{Q}_{kl}(t)x_l(t)+ Q_{kl}(t)v_l(t) +\dot{b}_k(t)$. The  velocity gradient between two frames can then be found as  $v'_{k,m}(t')=\dot{Q}_{kl}(t)Q_{ml}(t)+ Q_{kl}(t)Q_{mp}(t)v_{l,p}(t)$.

Therefore, the symmetric part of the velocity gradient between two frames is proven to be objective by $S'_{km}=\frac{1}{2}(v'_{k,m}(t')+ v'_{m,k}(t'))= Q_{kl}(t)Q_{mp}(t)\frac{1}{2}(v_{l,p}(t) + v_{p,l}(t))=Q_{kl}(t)Q_{mp}(t)S_{lp}$ where $\dot{Q}_{kl}(t)Q_{ml}(t)+ \dot{Q}_{ml}(t)Q_{kl}(t) =\frac{d}{dt}Q_{ml}Q_{kl}=\frac{d}{dt}\delta_{km}=0$.

Nevertheless, the antisymmetric part is found to be $\Omega'_{km}=\frac{1}{2}(v'_{k,m}(t') - v'_{m,k}(t'))= Q_{kl}(t)Q_{mp}(t)\Omega_{l,p} + \frac{1}{2}(\dot{Q}_{kl}(t)Q_{ml}(t) - \dot{Q}_{ml}(t)Q_{kl}(t))$.  If the rotation matrix $Q_{kl}$ is no longer time dependent,  ie.  $\dot{Q}_{kl}(t)Q_{ml}(t) = \dot{Q}_{ml}(t)Q_{kl}(t))=0$, $\Omega_{kl}$ is invariant. In other words, the antisymmetric part is Galilean invariant and only stays invariant between two frames with translation.

In MCT, the Cauchy stress is related  to the velocity gradient  and gyration through an objective strain-rate tensor, $a_{kl}=v_{l,k}+e_{lkm}\omega_m$ and $a'_{mn}=Q_{mk}Q_{nl}a_{kl}$. The objectivity of $a_{kl}$ can be proven through a process similar to the aforementioned paragraph on velocity gradient.  The orientation of inner structure is described by the director tensor, $\chi_{kK}$ (cf. eq. \ref{eq:director}).  The director and its time derivative between two frames with rigid body motions can be shown as
\begin{align}\label{eq:directorRate}
&\chi'_{kK}(t')=Q_{km}(t)\chi_{mK}(t)\nonumber \\
&\dot{\chi}'_{kK}=\dot{Q}_{km}\chi_{mK}+Q_{km}\dot{\chi}_{mK}\\
&e_{lkm}\omega'_m\chi'_{lK}=\dot{Q}_{km}\chi_{mK}+Q_{km}e_{amb}\omega_b\chi_{aK}\nonumber
\end{align}
where $\dot{\chi}_{mK}=e_{amb}\omega_b\chi_{aK}$, $\omega_b$ is the rotational velocity of an inner structure. After multiplying another director tensor on  eq. \ref{eq:directorRate} and utilizing the identity in eq. \ref{eq:inverse}, one can obtain
\begin{align}
    e_{mkp}\omega'_p=\dot{Q}_{kp}Q_{mp}+Q_{ma}Q_{kt}e_{atb}\omega_b
\end{align}

From the previous paragraph, one can recall the velocity  gradient described in two frames are  related as
\begin{align}
v'_{m,k}=\dot{Q}_{mp}Q_{kp}+Q_{ma}Q_{kt}v_{a,t}
\end{align}
Therefore, one can see that
\begin{align}\label{eq:A_objective}
 (v'_{m,k}+e_{mkp}\omega'_p)= Q_{ma}Q_{kt}(e_{etb}\omega_b+v_{a,t})
\end{align}
since $\frac{d}{dt}(Q_{kp}Q_{mp})=\frac{d}{dt}\delta_{km} =0$. Equation \ref{eq:A_objective} proves the strain rate tensor, $a_{km}$, is objective.

As opposed to using the velocity gradient in NS equations for vortex identifications with Q-criterion, MCT relies on the strain rate tensors. The classical Q-criterion with the velocity gradient can be found as the second invariant of the velocity gradient, i.e. $Q=\frac{1}{2}(v_{i,i}v_{j,j}-v_{i,j}v_{j,i})=\frac{1}{2}(\Omega_{ij}\Omega_{ji}-S_{ij}S_{ji})$, where $S_{ij}$ is the symmetric part and $\Omega_{ij}$ is the antisymmetric part of the velocity gradient. Following a similar derivation, the MCT strain rate tensor can also be divided into a sum of a symmetric and antisymmetric part.
\begin{align}
&S_{ij}^{\text{MCT}}=\frac{1}{2}(a_{ij}+a_{ji})=\frac{1}{2}(v_{j,i}+v_{i,j})\\
&\Omega_{ij}^{\text{MCT}}=\frac{1}{2}(a_{ij}-a_{ji})=\frac{1}{2}(v_{j,i}-v_{i,j}+2e_{jim}\omega_m)
\end{align}
It should be emphasized that since $a_{ij}$ is objective, the addition or subtraction between objective tensors, e.g. $S_{ij}^{\text{MCT}}$ and $\Omega_{ij}^{\text{MCT}}$, remain objective. As a results, an objective Q-criterion for MCT is proposed as the second invariant of the strain rate tensor, $a_{ij}$, ie.
\begin{align}
Q^{\text{MCT}}&=\frac{1}{2}(a_{ii}a_{jj}-a_{ij}a_{ji})\nonumber \\
&=\frac{1}{2}(v_{i,i}v_{j,j}-v_{j,i}v_{i,j}-2v_{j,i}e_{ijm}\omega_m+2\omega_m\omega_m)\nonumber\\
&=\frac{1}{2}(\Omega_{ij}^{\text{MCT}}\Omega_{ij}^{\text{MCT}}-S_{ij}^{\text{MCT}}S_{ij}^{\text{MCT}})
\end{align}

Using Cartesian Coordinate , the  objective Q-criterion can be written as
\begin{align}
Q^{\text{MCT}}=&v_{x,x}v_{y,y}+v_{x,x}v_{z,z}+v_{y,y}v_{z,z}\nonumber \\
&-(v_{x,y}v_{y,x}+v_{x,z}v_{z,x}+v_{y,z}v_{z,y})\nonumber\\
&-(v_{y,x}-v_{x,y})\omega_z-(v_{x,z}-v_{z,x})\omega_y\nonumber \\
&-(v_{z,y}-v_{y,z})\omega_x+\omega_x^2+\omega_y^2+\omega_z^2
\end{align}

\begin{figure}[h!]
\includegraphics[height=2 in]{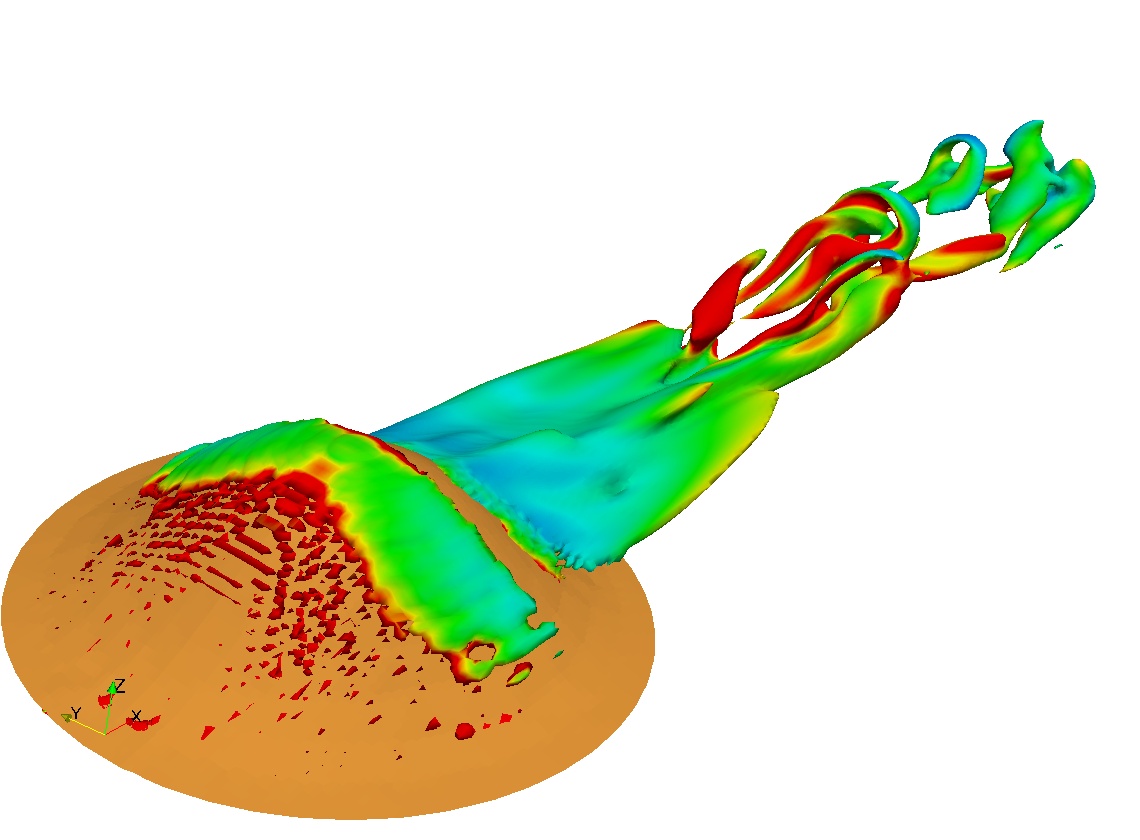}
\caption{Hairpin eddy structure identified by the objective Q-Criterion with MCT}
\label{fig:Q}
\end{figure}

The symmetric part is the same as the one in NS theory showing the normal expansion of the flow behaviors. However, the physical meaning of the anti-symmetric part, $\Omega_{ij}^{\text{MCT}}$, should be understood as absolute rotation. The off-diagonal part of an anti-symmetric matrix can be represented by a vector. Therefore, one can rewrite the antisymmetric part as a vector of absolute rotation ($\Omega^\text{AR}_k$), i.e.
\begin{align}
\Omega^\text{AR}_k&=e_{ijk}\Omega_{ij}^{\text{MCT}}\nonumber \\
&=e_{ijk}v_{j,i}-2\omega_k\nonumber \\
&\sim \nabla\times\vec{v}-2\vec{\omega}
\end{align}
The first half of the $\Omega^\text{AR}_k$ is vorticity ($\nabla\times\vec{v}$) describing the relative rotation between two inner structure while the second half ($\vec{\omega}$) is the self-spinning of an inner structure. In other words, $\Omega^\text{AR}_k$ measures the phase shift or the rotational speed difference between the relative rotation and the self-spinning motion. This is the true rotation between two inner structures in a continuum and it does not change even when observed from different reference frames. If $\Omega^\text{AR}_k$ is zero, it implies that the relative revolution between two inner structures is equal to the self-spinning motion. Therefore, two inner structures always face each other with the same side, like the Earth and the Moon. Without a global coordinate, the inner structure behaves as if there is no motion. Mathematically,  $\Omega^\text{AR}_k=0$ reduces MCT back to NS equations \cite{LopezChenPalochko2016}. This mathematical relation implies that if one believes vorticity can completely resolve all possible rotation without self-spinning gyration, NS theory and MCT are equivalent.

{{It is noted that Truesdell followed the momumental work by Grad \cite{Grad1952} and derived a balance law of internal rotation \cite{Truesdell1966,TruesdellRajagopal1999}. De Groot and Mazur also discussed a similar governing equation in their book \cite{DegrootMazur1962}. The concept of the internal rotation is similar to the new degrees of freedom, gyration, in MCT. However, De Groot and Mazur derived the balance law from a mechanics perspective so the time evolution of the intrinsic rotation is only governed by the antisymmetric part of Cauchy stress, i.e. $e_{kij}t_{ij}$ in eq. \ref{eq:MCT+AM}. On the other hand, the constitutive equation of gyration in MCT was derived from the classical nonequilibrium thermodynamics. Therefore, there is an additional moment stress, i.e. $m_{lk,l}$ in eq. \ref{eq:MCT+AM}. Consequently, there is a dissipation or diffusion mechanism in the balance law of angular momentum, i.e. $\omega_{k,ll}$ in eq. \ref{eq:MCT+AM_G}. The diffusion of gyration leads to the heat and eventually the irreversible entropy generation.}}

Figure \ref{fig:Q} shows the iso-surface of the objective Q-Criterion for the coherent eddy structure in the transonic flow over a three-dimensional bump. The iso-surface is colored by the magnitude of the absolute rotation ($\Omega^\text{AR}_k$). The hairpin structure of the  eddies are clearly seen  without being limited by the Galilean invariance.

\section{Conclusion}\label{sect:conclusion}
This work reviews the development of morphing continuum theory from both mathematical and physical perspectives. The complete MCT framework is derived under the framework of rational continuum mechanics for turbulence with subscale eddy structures. A second order finite volume method with second order shock preserving scheme is summarized along with the recent developments on the numerical methods for MCT.

A case of a transonic turbulence over a three-dimensional bump is compared among the MCT, NS and experiments. With less than 10\% of the mesh number required in NS-DNS, MCT-DNS provided a better prediction on the pressure coefficient and the pressure profile in the downstream turbulence. It shows that the multiscale MCT does not require the arbitrary fine mesh to resolve the subscale eddy motions. Instead, the subscale eddy motions are captured by the additional degrees of freedom, eg. gyration. 

In addition, MCT allows for an objective or frame-indifferent G-criterion for eddy or vortex identifications. The classical NS-based Q-criterion is only Galilean-invariant. It changes when the reference frames becomes time-dependent. The newly proposed MCT-based Q-criterion does not have this limitation and provides sounding results for coherent hairpin eddy structure in the supersonic turbulent flows over a bump.

Future works should be directed at investigating the energy transfer phenomena, shock structure and other essential characterizes in highly compressible turbulence with the multiscale framework of MCT and an affordable computational resources.

\section*{Acknowledgement}
This material is based upon work supported by the Air Force Office of Scientific Research under award number FA9550-17-1-0154.

\bibliography{ref}
\end{document}